\documentclass[prl,aps,twocolumn,floats,showpacspsfig]{revtex4}
\usepackage{amssymb}

\usepackage{epsfig}

\newcommand{\be}{\begin{equation}}
\newcommand{\ee}{\end{equation}}
\newcommand{\bea}{\begin{eqnarray}}
\newcommand{\eea}{\end{eqnarray}}

\newcommand{\p}{\partial}
\newcommand{\s}{\sigma}

\newcommand{\la}{\langle}
\newcommand{\ra}{\rangle}
\newcommand{\rd}{\mbox{d}}
\newcommand{\ri}{\mbox{i}}
\newcommand{\re}{\mbox{e}}

\begin{document}
\title{Friedel oscillations of Density of States  in a one-dimensional  Mott insulator and Incommensurate Charge Density Wave/Superconductor.}

\author{ A.  M. Tsvelik}
\affiliation{Department of  Condensed Matter Physics and Materials Science, Brookhaven National Laboratory, Upton, NY 11973-5000, USA}
\date{\today}

\begin{abstract}
Oscillations of local density of states generated by  a single  scalar  impurity potential are calculated for one-dimensional systems with dynamically generated charge or spin gap.  At zero temperature the oscillations develop at finite wave vector ($\pi$ for the Mott insulator and $2k_F$ for ICDW/SC) and at frequencies larger than the soliton spectral gap $m$. Their amplitude has a broad maximum at $\omega \approx 3m$, where $m$ is the gap magnitude.  
\end{abstract}

\maketitle
 Modern Single Electron Tunneling Microscopy (STM) techniques provide a plephora of data by measuring  minute details of  coordinate dependence of the differential conductance $\s({\bf x}) = \rd I/\rd V$. An especially rich picture emerges in the presence of impurities which create Friedel oscillations in $\s({\bf x})$. An ultimate goal for a theory is to decipher the interference pattern of these  oscillations and extract from it information about the spectrum and interactions in the underlying system. Since the  conductance is directly related to the single electron density of states (DOE), theory should aim at calculation of this quantity:
\bea
&& \s(V, {\bf x}) = C({\bf x})\rho(\omega = eV, {\bf x}); \nonumber\\
&& \rho(\omega, {\bf x}) = \frac{1}{\pi}\Im m G^{(R)}(\omega; {\bf x},{\bf x})
\eea
where $C({\bf x})$ is an {\it a priori} unknown matrix element. 

 In majority  of  systems studied by now by STM the disorder is  sufficiently weak  for momentum to  remain a good quantum number. For instance, observing  sharp peaks in the  Fourier images of   the STM measurements in high-T$_c$ superconductors, like in \cite{davis},  one may only conclude that the impurities are sufficiently far apart so  that the Friedel oscillations from different impurities has room to develop without  quenching each other. Therefore a reasonable starting point for the theory would be to study the Friedel oscillations of DOS from a single impurity. For a $d$-wave superconductor this was done in \cite{dung} using BCS theory combined with a $T$-matrix approximation. The results are  now widely used  for the analysis of STM experiments on high-T$_c$ superconductors giving  very reasonable answers for the quasiparticle spectrum not far from the nodal points. Needless to say that since the calculation is based on a picture of well-defined weakly interacting quasiparticles, it cannot hold close to the antinodal point where the STM measurements also yield very rich structure \cite{Davis2} which requires a careful analysis.

 In general, Friedel oscillations of DOS are not very simple phenomenon and allow a straighforward interpretation only when there are well defined quasiparticles in the system. In the first order of perturbation theory in the impurity potential $V$ one has
\begin{widetext}
\bea
&& \delta\rho(\omega;{\bf r}-{\bf r}_0) = \sum_n\frac{\re^{-\beta E_n} + \re^{-\beta E_m}}{Z}\delta(\omega + E_n - E_m)(A_{nm} + B_{nm})\label{pert}\\
&& A_{nm} = \sum_{m\neq k}\left\{\frac{\exp[\ri{\bf P}_{mk}({\bf r}-{\bf r}_0)]}{E_m - E_k}\la k|V(0)|m\ra\la m|\hat\psi^+(0)|n\ra \la n|\hat\psi(0)|k\ra + c.c.\right\}\\
&& B_{nm} = \sum_{k\neq n}\left\{\frac{\exp[\ri{\bf P}_{nk}({\bf r}-{\bf r}_0)]}{E_n - E_k}\la n|V(0)|k\ra\la k|\hat\psi(0)|m\ra \la m|\hat\psi^+(0)|n\ra + c.c.\right\} \nonumber
\eea
\end{widetext}
where $Z$ is the partition function, $|n\ra,|m\ra, |k\ra$, $E_{n,m,k}$ and ${\bf P}_{n,m,k}$ are eigenfunctions and energy and momentum eigenvalues of the many-body Hamiltonian governing the system.$\hat\psi^+$, $\hat\psi$ are electron 
creation and annihilation operators. All operators are taken at the same spacial point 0. As follows from these expressions, the relation between the Friedel oscillations and the excitation spectrum is not that straighforward. The situation is simplified only at small temperatures and  when the many-body states in question can be approximated as quasiparticle ones. To see that let us consider $\omega > 0$. At $\beta^{-1} \equiv T=0$ $|n\ra = |0\ra$ is the ground state and $E_m = \omega$. Taking the Fourier transform of (\ref{pert}) in real space one fixes ${\bf P}_{km}$ to be the external wave vector. Fixing ${\bf P}_{km}$ may fix $E_k$ and lead to singularity in the Friedel oscillation amplitude, but only in one-dimension and only if the eigenenergies in question belong to quasiparticles. In $D > 1$ equation $E_m = \omega$ determines a surface in momentum space and the singularity is smeared by integration along this surface. In the same way the singularities are weakened when the eigenenergies $E_{n,k}$ belong to continuum of states which cannot be parametrized by a single momentum.  Therefore one is driven to the conclusion that in order to decipher the interference pattern concrete and model dependent expressions for DOS are necessary.   
 
  In this paper I describe Friedel oscillations of local DOS for two strongly correlated systems - one-dimensional Mott insulator and one-dimensional Incommensurate Charge Density Wave (ICDW)/Superconductor. The motivation behind the calculations is twofold. First, these are strongly correlated systems which require nonperturbative approach. Nonperturbative results in this area predominantly concern Tomonaga-Luttiger liquids (see \cite{fradkin}). In the systems of choice where spectral gaps are generated dynamically and  the long lange order is destroyed by gapless fluctuations the only available result is \cite{dirk}. Second, in my previous publication with Chubukov \cite{phen} we suggested that the dynamics of  the antinodal regions in high-T$_c$ superconductors is essentially one-dimensional. Therefore  these calculations may be even relevant to the high-$T_c$ problem after all.   

 I will study the case when the spectral gap (be it the charge gap in  Mott insulator or the spin gap in ICDW/Superconductor) is small compared to the bandwidth. The chemical potential is in the middle of the gap. The relative smallness of the gap enables me to  employ field theory methods and bosonization. These methods together with the background information on the models in question are described in may review articles and books, in particular in \cite{book}.  In the continuum limit both the   Mott insulator and ICDW/SC are described by a universal Hamiltonian. The corresponding Hamiltonain density is a sum of two commuting parts governing  dynamics of the charge and the spin collective modes. For the Mott insulator we have  
\bea
&& {\cal H} = {\cal H}_c + {\cal H}_s\\
&& {\cal H}_c = \frac{v_c}{2}[K_c(\p_x\theta_c)^2 + K_c^{-1}(\p_x\phi_c)^2]\nonumber\\
&&  - \mu\cos(\sqrt{8\pi}\phi_c)\label{charge}\\
&& {\cal H}_s = \frac{v_s}{2}[K_s(\p_x\theta_s)^2 + K_s^{-1}(\p_x\phi_s)^2]
\eea
where $K_c < 1, K_s$ and $v_c,v_s$ are the Luttinger parameters  and the velocities in the  charge and spin sector respectively and $\mu > 0$ is the coupling constant. Fields $\phi_a,\theta_a$ obey the standard commutation relations:
\bea
[\theta_a(x),\phi_b(y)] = -\ri\delta_{ab}\Theta_H(x-y)
\eea
 where $\Theta_H(x)$ is the Heaviside function. For ICDW/SC one has to interchange  charge and spin indices. Then $K_s$ becomes  the Luttinger parameter in the charge channel.  For $1/2 < K_s < 2$ both $2k_F$ charge susceptibility and pairing susceptibility are singular at zero frequency and temperature diverging as 
\be
\chi(2k_F,\omega =0) \sim T^{-2 + K_s}, ~~ \chi_{pair}(\omega=0) \sim T^{-2 + 1/K_s}
\ee
What order will eventually emerge when such one-dimensional systems are coupled together depends on whether $K_s$ greater or smaller than 1 and to some extent on the strength of the interchain interactions.  Since both CDW and SC channels are singular, I do not distinguish between the two and label the model ICDW/SC. However, to avoid confusion from now on I will discuss only the Mott insulator reserving ICDW/SC system  for the final discussion. 

 Model (\ref{charge}) is the sine-Gordon model; it is well studied  
 and  plenty of exact results are available, including results for its correlation functions. At $K_c < 1$ the spectrum has gap(s). The excitations always include solitons and antisolitons carrying electric charge $\pm e$ (for ICDW it would be spin 1/2) and for $K_c < 1/2$ also their neutral bound states (excitons). The asymptotics of the single electron Green's function is dominated by single soliton emission processes \cite{ZamLuk}.
 
All spectral gaps are of the same order. The Mott gap is 
\be
m \sim \Lambda (\mu/\Lambda)^{1/2(1-K_c)}
\ee
where $\Lambda$ is the ultraviolet cut-off and is probably of order of the bandwidth. 

 For a clean Mott insulator DOS is featureless (apart from the spectral gap) and is given by ( see Eqs.(9,10) for $N=2$ in \cite{Ess}):
\bea
\rho(\omega) \sim  (|\omega| - m)^{2\gamma}, ~~ \gamma = \frac{1}{4}(\sqrt{K_s} - 1/\sqrt{K_s})^2 
\eea
This is in contrast with band insulator; for Mott insulator with spin 1/2 electrons quantum fluctuations wipe out the singularity at the gap. We will see that impurities change this strongly enhancing the amplitude of  Friedel oscillations at frequencies of order of $m$. 

Let us consider a single non-magnetic impurity coupled to the total density operator and placed at point $x_0$. 
The most relevant contribution to the Hamiltonian density comes from  the  backward scattering and is described by the operator
\bea
V = U(2k_F)\cos[\sqrt{2\pi}\phi_c(x_0)]\cos[\sqrt{2\pi}\phi_s(x_0)] \label{Vop}
\eea
where $U(2k_F)$ is proportional to the matrix element of the impurity potential.  In what follows I will replace  $\cos(\sqrt{2\pi }\phi_c)$ by its vacuum expectation value. By doing so I neglect scattering of the massive solitons on fluctuations of the potential.  It is justified when the bare scattering potential is weak $U(2k_F) << \Lambda$. The alternative approach taken in \cite{dirk} was to replace the entire potential (\ref{Vop}) by a solid wall which is fine if the bare impurity scattering is close to the unitary limit and, in my opinion, substantially overestimates the contribution from  such scattering when the potential is weak. After the above replacement  the impurity scattering acts only in the spin sector where we have the boundary sine-Gordon problem:
\bea
&& {\cal H}_s = \frac{v_s}{2}[K_s(\p_x\theta_s)^2 + K_s^{-1}(\p_x\phi_s)^2] - \nonumber\\
&& \lambda\delta(x-x_0)\cos[\sqrt{2\pi}\phi_s(x_0)] \label{spin}
\eea
where $\lambda = U(2k_F)\la \cos[\sqrt{2\pi}\phi_c]\ra  \sim U(2k_F)(m/\Lambda)^{K_c/2}$. This model is exactly solvable \cite{Goshal} and at the particular value of the coupling $K_s =1$ can be even reduced to the model of free fermions \cite{guinea}.   Another free fermion point (a trivial one) is $K_s =2$.  However, calculation of correlation functions of bosonic exponents is not an easy problem and in its entirety has not yet been solved, even at $K_s =1$. Some help comes from the fact that at $K_s < 2$ the impurity scattering potential scales  to strong coupling and below the energy scale 
\be
E^* \sim \Lambda(\lambda/\Lambda)^{2/(2- K_s)} \label{Ec}
\ee
 can be replaced by Dirichlet condition 
\be
\phi_s(x_0) =0 \label{BC}
\ee
 This gives an easy way  to calculate asymptotics of the correlation functions of the bosonic exponents using the method of images. In order to get a feeling for the magnitude of errors originating  from deviations from the asymptotic regime, we will consider Friedel oscillations of the particle density 
\bea
&& \la \rho(x) - \rho_0\ra = \la \cos(2k_F x + \sqrt{2\pi}\phi_c)\cos(\sqrt{2\pi}\phi_s)\ra \nonumber\\
&& \approx \cos(2k_Fx) m^{K_c/2}\la \cos[\sqrt{2\pi}\phi_s(x)]\ra \label{dens}
\eea
where the charge cosine was replaced by its vacuum expectation value. At $K_s=1$ the average in the right hand side of (\ref{dens}) was calculated in \cite{fisher} (see Eq.(3.12)).Taking the Fourier transform of this formula  we obtain
\bea
\la \rho(2k_F + q)\ra \sim \frac{\left(1 + \sqrt{1 + (E^*/q)^2}\right)^{1/2}}{\sqrt{q^2 + {E^*}^2}}
\eea
One can check that a relative deviation of this function from its low-$q$ asymptotic value $|q|^{-1/2}$ exceeds 20 percent only at $(q/E^*) > 2.5$. 

  Now let us come back to our original task: calculation of oscillations of DOS. The bosonized expression for the fermionic operator is
\bea
&& \psi_{\s}(x) = \re^{-\ri k_F x}R_{\s}(x) + \re^{\ri k_F x}L_{\s}(x)\\
&& R_{\s} = \frac{\eta_{\s}}{\pi a_0}\re^{\ri\sqrt{\pi/2}(\phi_c + \theta_c)}\re^{\s\ri\sqrt{\pi/2}(\phi_s + \theta_s)}\nonumber\\
&& L_{\s} = \frac{\eta_{\s}}{\pi a_0}\re^{\ri\sqrt{\pi/2}(-\phi_c + \theta_c)}\re^{\s\ri\sqrt{\pi/2}(-\phi_s + \theta_s)}\label{R,L}
\eea
where $\eta_{\s}$ are the Majorana zero modes (Klein factors), $a_0$ is the lattice constant and $\s = \pm 1$.
 
The single-particle density of states is related to the single-particle Green's function at coinciding spacial points:
\bea
&& G(\tau;x,x) = G_{smooth} + G_{oscil}\\
&& G_{smooth} = \la\la R(\tau,x) R^+(0,x)\ra\ra + \nonumber\\
&& \la\la L(\tau,x) L^+(0,x)\ra\ra\\
&& G_{oscil} = \left[\re^{2\ri k_F x}\la\la L^+(\tau,x) R(0,x)\ra\ra + H.c.\right] \label{osc}
\eea
where $\tau$ is Matsubara time. Let the reader note that $G_{oscil} =0$ in a clean sample. 

The operators $R,L$ and their Hermitian conjugate factorize into charge and spin parts (\ref{R,L}) governed by different Hamiltonians (\ref{charge}, \ref{spin}) respectively. Since the impurity contributes mostly to the spin Hamiltonian, for the charge sector one can use the results obtained in \cite{G}. Then  for the oscillatory oscillatory part (\ref{osc}) of the Green's function we obtain
\bea
&& G_{oscil} =\alpha m^{1/2}K_0(m\tau)S(\tau,x-x_0)\label{osc1}\\
&& S(\tau,x-x_0) = \label{S}\\
&& \left[\re^{2\ri k_F x}\la\la \re^{\ri\sqrt{\pi/2}(\phi_s + \theta_s)(\tau,x)}\re^{\ri\sqrt{\pi/2}(\phi_s - \theta_s)(0,x)}\ra\ra + c.c.\right]\nonumber
\eea
where $\alpha$ is a numerical coefficient. The $K_0$ function in (\ref{osc1}) comes from the charge sector (the corresponding correlator was calculated in \cite{G}). Therefore the  fact that in the clean system object (\ref{osc})  vanishes altogether is due to  the spin sector remaining  critical. The latter means  that its right and left  sectors  do not couple to each other and $S =0$.  The scattering from the impurity(ies) connects the right and left sectors and  $S(\tau,x)$ (\ref{S}) is no longer zero. Thus impurities reveal the parity violation which exists in the Mott insulator already in the clean case, but is not directly observable for a lack of a suitable local operator. 

 Function (\ref{S}) can be represented as 
\bea
S(\tau,x) = \frac{1}{|x|^{2d}}{\cal F}(\frac{v_s\tau}{x}, xE^*/v_s) \label{S1}
\eea
where $d = (K + 1/K)/8$ and ${\cal F}(y,z)$ is a scaling function.
In the strong coupling limit (that is at the distances or times much larger than $1/[E^*]$) the impurity is substituted by the condition (\ref{BC}). This condition is equivalent to  the conditions
\be
\phi_s(x -x_0) = -\phi_s(x_0 -x), ~~ \theta_s(x-x_0) = \theta(x_0 -x)
\ee
 Then  correlation functions of bosonic exponents are calculated by the method of images, like in electrostatics.  As a result in the strong coupling limit we obtain universal (that is independent of the bare impurity potential) asymptotics:  
\bea
&& G_{smooth} = \frac{Z \re^{-m|\tau|}}{2\pi \tau}|\tau|^{-\gamma}\left(1 + \frac{(\tau v_s)^2}{4(x-x_0)^2}\right)^{K_s/16}\\
&& G_{oscil} = \alpha \cos[2k_F(x - x_0)] \times\label{oscil}\\
&& \frac{ m^{1/2}K_0(m\tau)|\tau|^{(K_s - 1/K_s)/4}}{|(x -x_0)/v_s|^{K_s/4}[\tau^2 + 4(x - x_0)^2/v_s^2]^{K_s/8}}\nonumber
\eea
where factor $Z$ was calculated in \cite{ZamLuk}  and $\alpha \sim 1$. We perform a double Fourier transformation of (\ref{oscil}), both in space and Matsubara time and then perform the necessary analytic continuation $\ri\omega \rightarrow \omega + \ri 0$. The answer can be written as 
\bea
&& \delta\rho(\omega >0; 2k_F + q) \sim \\
&& |q|^{-1 + K_s/2} \int_0^{2(\omega -m)/v_s|q|}\rd y A(\omega -y|q|v_s/2)f(y)\nonumber\\
&& f(y) = \int_{-y}^y \rd z (y^2 -z^2)^{-1 + K_s/8}|1 + z|^{-1 + K_s/4} 
\eea
Here 
\bea
&& A(\omega) = [(\omega/m)^2 -1]^{-1/2 -a}(\omega/m)^{a/2}\times\nonumber\\
&& F\left(- a/2, 1/2 - a/2; 1/2 -a; 1- (m/\omega)^2\right)
\eea
where $a = (K_s - 1/K_s)/4$ and $F(a,b;c;x)$ is the hypergeometric function. Numerical evaluation of these integrals show that that at $1/2 < K_s <2$ the Friedel oscillations do not disperse and can be approximated by 
\bea
\delta\rho(\omega,q +2k_F) \sim |q|^{-2 +K_s/2}{\cal G}(\omega/m -1;K_s)
\eea
where function ${\cal G}(x;K_s)$ is depicted on Figs.1,2 for $K_s =1,2$. 


\begin{figure}[ht]
\begin{center}
\epsfxsize=0.3\textwidth
\epsfbox{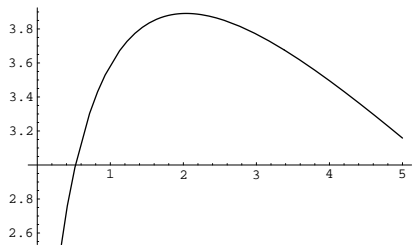}
\end{center}
\caption{Function ${\cal G}(x;1)$.}
\label{fig:K1}
\end{figure}

\begin{figure}[ht]
\begin{center}
\epsfxsize=0.3\textwidth
\epsfbox{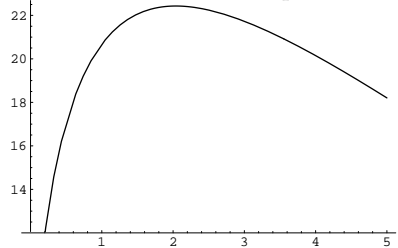}
\end{center}
\caption{Function ${\cal G}(x;2)$.}
\label{fig:K2}
\end{figure} 

 Thus we have found that impurities in 1D Mott insulator (or ICDW/SC) produce $2k_F $ Friedel oscillations of DOS. These oscillations occur at frequencies larger than the Mott gap (the soliton gap) and at $T=0$ their  amplitude has a maximum at $\omega \approx 3m$ and a strong singularity at $2k_F$. Near  the singularity the  Fourier transform of the DOS is essentially dispersionless.  In our calculations DOS (including the  Friedel oscillations) is  even in frequency. This is due to the fact that the electron spectrum is linearized and the chemical potential is exactly in the middle of the gap. The latter feature can be trivially corrected by a chemical potential shift.
 
For Mott insulator with SU(2) symmetry in the spin sector $K_s =1$ and $a=0$. For ICDW/SC spin and charge sectors $s$ and $c$ are interchanged; then $K_s$ corresponds to the Luttinger parameter in the charge sector and its value is not fixed by any symmetry. The phenomenology developed for the cuprates in \cite{phen} suggests  $K_s \approx 2$. Note that at $K_s =2$  operator (\ref{Vop}) becomes marginal and $E^*$ increases exponentially when the value of $K_s$ approaches 2 from below.   


I am  grateful to J.C. Davis, F. H. L. Essler, D. Schuricht and A. Chubukov for valueable discussions and acknowledge the support from
 US DOE under contract number DE-AC02 -98 CH 10886.

\end{document}